# Light storage based on four-wave mixing and electromagnetically induced transparency in cold atoms

Jinghui Wu[1], Yang Liu[1], Dong-Sheng Ding[1], Zhi-Yuan Zhou, Bao-Sen Shi[†], and Guang-Can Guo

*Key Laboratory of Quantum Information, University of Science and Technology of China, Hefei 230026, China*

[†]*drshi@ustc.edu.cn*

**Abstract:** We performed an experiment to observe the storages of an input probe field and an idler field generated through an off-axis four-wave mixing (FWM) process via a double-lambda configuration in a cold atomic ensemble. We analyzed the underlying physics in detail and found that the retrieved idler field came from two parts if there was no single-photon detuning for pump pulse: part 1 was from the collective atomic spin (the input probe field, the coupling field and the pump field combined to generate the idle field through FWM, then the idler was stored through electromagnetically induced transparency.); part 2 was from the generated new FWM process during the retrieval process (the retrieved probe field, the coupling field and the pump field combined to generate a new FWM signal). If there was single-photon detuning for pump pulse, then the retrieved idler was mainly from part 2. The retrieved two fields exhibited damped oscillations with the same oscillatory period when a homogeneous external magnetic field was applied, which was caused by the Larmor spin precession. We also experimentally realized the storage and retrieval of an image of light using FWM for the first time. In which, an image was added into the input signal. After the storage, the retrieved idler beams and input signal carried the same image. This image storage technique holds promises for application in image processing, remote sensing and quantum communication.

PACS number(s): 42.30.-d, 42.50.Gy, 42.50.Md, 42.70.Ln

## 1. Introduction

Electromagnetically induced transparency (EIT) [1, 2] can be used to make a resonant, opaque medium transparent by means of quantum interference. Meanwhile, the optical properties of the medium will be dramatically modified, e.g. great enhancement of nonlinear susceptibility in the spectral region of induced transparency and associated steep dispersion. The steep dispersion is necessary to realize light-slowing and -storage. In a typical EIT system, the group velocity of probe pulse could be reduced to zero by adiabatically switching off the coupling field and the probe pulse is subsequently stored as a collective atomic spin excitation, which enables a quantum memory for light [3-5]. A number of progresses have been made in many candidates of quantum memory, such as a cold atomic ensemble trapped in a magneto-optical trap (MOT) [6], a warm atomic cloud in vapor cell [3], the Bose-Einstein condensate [2], the rare-earth-doped crystals at low temperatures [7], etc. An optically thick atomic ensemble is considered as a promising candidate for quantum memory [1-6, 8, 9]. A scalable quantum repeater protocol based on atomic ensembles and linear optics (DLCZ scheme) has been proposed [10] in 2001. Since then, a number of works have reported experimental studies of generation and subsequent retrieval of DLCZ collective excitations [11-18].

EIT-based quantum memory could be understood in term of a quasi-particle, a so-called dark state polariton (DSP) [5]. A DSP consists of a photonic component and a hyperfine spin wave (a collective matter excitation) component, in which the ratio between the light and matter components can be controlled by changing the amplitude of a control laser field. The storage and subsequent retrieval of a signal field can be achieved by the extinction and subsequent reactivation of the control field after a given storage time. Another technique for storing a light is based on delayed four-wave mixing (FWM) process [19], where the generated FWM field can be delayed and stored in atomic spin excitation. Besides, collapse and revival of DSP in an external uniform magnetic field has been experimentally and theoretically studied in Ref. [20-22]. The DSP collapses in a doped crystal ($Pr^{3+}$:$Y_2SiO_5$) has also been observed and it could be controlled by tuning the magnitude of the applied external magnetic field [23]. The enhanced nonlinear susceptibility is useful in frequency mixing process, such as FWM [19]. Combining the EIT and FWM, the storage of light with a uniform spatial profile using FWM has been studied [19] in a hot atomic rubidium vapor, where the input signal and the generated idler are simultaneously stored by adiabatically switching off the pump laser. This kind of storage technique could improve the fidelity of the signal stored for practical usage [24] by capturing the whole input signal pulse and transferring it into the atomic coherence, which could not be achieved by traditional EIT storage technique. Therefore this storage technique needs and also is worth further study for understanding the underlying mechanism in detail.

Besides, so far, there is no any report on the storage and retrieval of a real image using FWM both in a cold and a hot atomic ensemble. Usually, one-dimensional optical information is stored, such as light pulses with a uniform spatial profile (i.e., no spatial information) and variation temporally only. However, the research interests have been recently extended to high-dimensional information storage (specially the spatial information), such as orbital angular momentum [25], transverse momentum and position [26], multiple transverse modes [27-29], even the ghost images [30], etc. A high-dimensional state shows some interesting properties compared with a one-dimensional state. For example, high-dimensional entangled states enable us to achieve more efficient quantum information processing [31].

In this work we experimentally study the storages of an input probe field and an idler field generated through an off-axis FWM process via a double-lambda configuration in a cold atomic ensemble. In our experiment, a pump pulse, a coupling pulse and a probe pulse combine to generate an idler pulse. We want to store the probe and the idler together. The coupling and probe pulse are resonant with atomic transitions. We find that the retrieved idler signal consists of two parts if the pump pulse has no single-photon detuning: part 1 comes from the collective atomic spin wave, which is established by the process: the input probe field, the coupling field and the pump field are combined to generate the idler field through FWM, then the idler is stored through EIT; part 2 comes from another FWM process occurring during the retrieval process (the retrieved probe field, the coupling field and the pump field combine to generate a new FWM signal). We experimentally find that when the single-photon detuning of pump is very large, the idler from part 1 is difficult to be stored, therefore the retrieved idler after the storage is mainly from part 2. On the contrary, if the pump field is resonant to atomic transition, then part 1 could also be stored by EIT, the observed idler after storage is the combination of two parts. Our results are identical somehow to the report with a hot atomic vapor in Ref. [32], where a seeded idler is input along with a probe pulse. It is found there that the seeded idler is not independently preserved during the storage process. These results are definitely useful for further understanding the mechanism of this storage technique.

We also observe beat interference between the retrieved probe and idler (two signals have the frequency difference of 80MHz). In the presence of an external uniform magnetic field, the retrieval efficiencies of two pulses exhibit damped oscillations with a period of 2.7μs, which could be interpreted by the Larmor precession caused by an applied external magnetic field.

Besides, we also experimentally realize the storage and retrieval of an image imprinted on the input signal using FWM for the first time. We find that both retrieved input signal and the generated idler beams in the FWM process carry the same image. This image storage technique holds promises for application in image processing, remote sensing and quantum communication.

## 2. Simplified theoretical analysis

For the experiment, we use the Zeeman sublevels of the degenerate two-level system associated with the $D_1$ transition of $^{85}$Rb atoms ($|5^2S_{1/2}, F=3> \rightarrow |5^2P_{1/2}, F=2>$) (Fig. 1). The repetition of the experiment is 34Hz. During each experimental period, the atoms are loaded for 29.2ms and the experimental window is 700μs. The cooling beams and Ioffe coils are shuttered off during the experiment. The repumper is always on to keep the atoms in the state $|5^2S_{1/2}, F=3>$. Coupling laser resonantly couples the transition $|5^2S_{1/2}, F=3> \rightarrow |5^2P_{1/2}, F=2>$ with σ- polarization and the probe field has the same frequency but opposite circular polarization with the coupling beam. The strong coupling and weak probe form a generic Λ-type EIT configuration. The strong pump is σ+ polarized and is set to be red detuning of 80MHz with respect to $D_1$ transition. Here, we define $\omega_m$ and $\Delta_m$ ($m=pr$, $p$, $c$ and $i$) as the frequency and detuning of probe, pump, coupling and idler fields respectively, and $\omega_{ij}$ as the energy difference between states $|i>$ and $|j>$. So we have $\Delta_{pr}=\omega_{31}-\omega_{pr}$, $\Delta_p=\omega_{31}-\omega_p$, $\Delta_c=\omega_{32}-\omega_c$, $\Delta_i=\omega_{32}-\omega_i$. These three lasers compose a double-Λ FWM level system. The coupling and the pump are switched on before signal laser by ~20μs for the preparation of initial states. Hence, the atoms are appropriately pumped into the states $|5^2S_{1/2}, F=3, m_F=-3>$ and $|m_F=-2>$. Therefore, there are two independent FWM cycles (shown in Fig. 1). Because of the independence of these two cycles, we will take one cycle into account below. The transparent window for the probe pulse induced by the control field is accompanied by a reduced group velocity. Similarly the atom ensemble is transparent for the generated idler field in the presence of the pump beam, which is σ+ polarized and 80MHz red detuned to the transition of $|5^2S_{1/2}, F=3 >$ to $|5^2P_{1/2}, F=2>$.

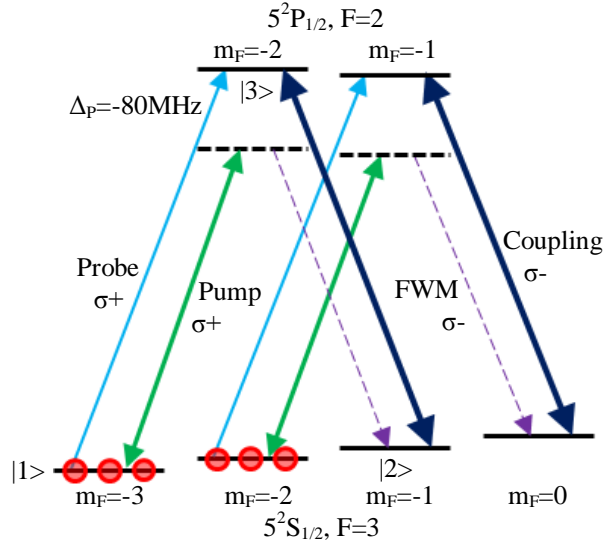

Fig. 1 (Color online) The $D_1$ transition of $^{85}$Rb creates a double-$\Lambda$ configuration. Coupling is resonantly connected to transition $|2\rangle$ ($5^2S_{1/2}$, $F=3$, $m_F=-1$) $\rightarrow |3\rangle$ ($5^2P_{1/2}$, $F=2$, $m_F=-2$), and probe is to state $|1\rangle$ ($5^2S_{1/2}$, $F=3$, $m_F=-3$) to state $|3\rangle$. The pump is detuned by 80MHz to the red of the optical transition connecting $|1\rangle$ and $|3\rangle$.

The time-dependent interaction Hamiltonian for this system in a rotating frame is

$$H_{int} = -\frac{\hbar}{2}\left[(\Omega_{pr}e^{i\Delta_{pr}t} + \Omega_p e^{i\Delta_p t})\sigma_{31} + (\Omega_i e^{i\Delta_i t} + \Omega_c e^{i\Delta_c t})\sigma_{32} + H.c.\right], \quad (1)$$

where $\Omega_{pr}(t)$, $\Omega_p(t)$, $\Omega_c(t)$ and $\Omega_i(t)$ denote the Rabi frequencies of probe, pump, coupling and idler fields respectively. According to energy conservation law, we have $\Delta_p + \Delta_c = \Delta_s + \Delta_i$. We define $\Delta_1 = \Delta_{pr}$, $\Delta_2 = \Delta_{pr} - \Delta_c$ and $\Delta_3 = \Delta_{pr} - \Delta_c + \Delta_i = \Delta_p$ as one-, two-, and three-photon detuning, respectively. The dynamics of laser-driven atomic system is governed by master equation for the atomic density operator:

$$\frac{d\rho}{dt} = \frac{1}{i\hbar}[H_{int}, \rho] - D. \quad (2)$$

$D$ is the decoherence matrix:

$$D = \begin{bmatrix} -\Gamma_{31}\rho_{33} & \frac{\gamma_2 \rho_{12}}{2} & \frac{\gamma_{31}\rho_{13}}{2} \\ \frac{\gamma_2 \rho_{21}}{2} & -\Gamma_{32}\rho_{33} & \frac{\gamma_{32}\rho_{23}}{2} \\ \frac{\gamma_{31}\rho_{31}}{2} & \frac{\gamma_{32}\rho_{32}}{2} & \Gamma_3 \rho_{33} \end{bmatrix}, \quad (3)$$

where $\Gamma_{31}$ and $\Gamma_{32}$ are the spontaneous emission rates from state $|3\rangle$ to states $|1\rangle$ and $|2\rangle$, respectively. We have also introduced dephasing processes with rates of $\gamma_3$ and $\gamma_2$. For convenience, we define the total spontaneous emission rate out of state $|3\rangle$ as $\Gamma_3 = \Gamma_{31} + \Gamma_{32}$. The coherence decay rates are defined as $\gamma_{31} = \Gamma_3 + \gamma_3$, $\gamma_{32} = \Gamma_3 + \gamma_3 + \gamma_2$, and $\gamma_{21} = \gamma_2$. Assuming that $\rho_{11} \approx 1$ and $\rho_{22} = \rho_{33} \approx 0$, the steady-state solution for $\rho_{12}(t)$ is

$$\rho_{12} = \frac{\Omega_{pr}\Omega_c^*}{4\Delta_1\Delta_2 - |\Omega_c|^2} + \frac{\Omega_p\Omega_i^*}{4\Delta_2\Delta_3 - |\Omega_p|^2 \Delta_3/\Delta_1}, \tag{4}$$

where $\Delta_1 = \Delta_{pr} + i\gamma_{31}/2$, $\Delta_2 = \Delta_{pr} - \Delta_c + i\gamma_2/2$ and $\Delta_3 = \Delta_p + i\gamma_{31}/2$ are the complex detunings. The expression of $\rho_{12}$ reflects the coherence between ground states, depends on a combination of the probe and idler optical fields. It relates to two different EIT processes: the probe-coupling and the idler-pump respectively. Thus we could see that both probe and idler pulses could be stored in a long-lived spin excitation by turning off the coupling and pump fields simultaneously. If the detuning $\Delta_p$ is large (i.e., $\Delta_3$ is large), the second term of Eq. (4) is close to zero, then the idler field cannot be stored in the atomic spin excitation.

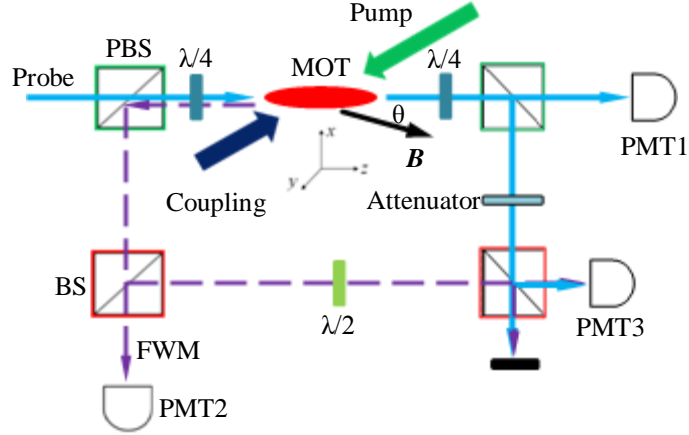

Fig. 2 (Color online) Simplified experimental setup. The photomultiplier tubes PMT1 and PMT2 (H10721, Hamamatsu) detect the signals. Meanwhile the beating of the two pulses is detected by PMT3. An attenuator is used in the probe channel to match the intensity of the two retrieved pulses so as to have a good visibility of the interference fringes.

## 3. Storage of light with a uniform spatial profile

We build up our experimental setup using an off-axis, counter-propagating geometry that introduced in Ref. [33]. The three input beams come from the same external cavity diode laser (ECDL, Toptica100) with a wavelength of 795nm in order to keep the phase coherence among them. The laser is locked resonantly to the transition $|5^2S_{1/2}, F=3\rangle$ to $|5^2P_{1/2}, F=2\rangle$ and is divided into three beams by two beam splitters. All the beams pass through respective acousto-optic modulators (AOMs) for beam switching and frequency shifting. Three single-mode optical fibers direct them to a cigar-shaped atomic cloud prepared in a two-dimensional MOT [34]. The atom ensemble has an on-resonance optical depth of 38. As shown in Fig.2, the coupling and pump beams, with diameters of 3.2mm and 2.2mm respectively, counter-propagate with each other and completely overlap in the atomic cloud. The probe beam propagates along the long axis of the atomic cloud and is focused to the center of the cloud. The full pulse width at half maximum of the probe is set to be 1.3μs. The generated idler signal is detected in the opposite direction with respect to the probe beam according to the phase-matching condition of $\vec{k}_{probe} + \vec{k}_{coupling} + \vec{k}_{pump} + \vec{k}_{idler} = 0$, where $\vec{k}$ is the wave vector. The coupling-pump axis has an angle of two degrees with respect to the probe-idler axis.

When the condition of FWM is satisfied, the input of a weak probe pulse with a Gaussian shape will be accompanied by a generated idler signal. When we switch off the coupling and pump fields adiabatically, both the probe and idler pulses will be stored in the atomic ground-state coherence $\rho_{12}(z, t)$.

$$\rho_{12}(z,t) = [g_{pr}E_{pr}(z,t) + g_i E_i(z,t)]\exp(-\frac{t}{2\tau}), \qquad (5)$$

where $g_{pr}$ and $g_i$ are the nonlinear coupling coefficients related to probe and idler respectively. $E_{pr}(z, t)$ and $E_i(z, t)$ are the time- and position-dependent amplitudes of probe and idler. The time constant $\tau$ shows an exponential decay of $\rho_{12}$, which is estimated to be around 32μs according to our measurement. In this experiment, because the single-photon detuning of the pump is quite large, the idler field is difficult to be stored in the atomic coherence, and most part of it leaks through atomic ensemble directly, so $\rho_{12}$ is mainly contributed from the probe. The storage of the idler here is similar to the storage scheme based on a far off-resonant two-photon transition, i.e., far off-resonant Raman memory [35] with a hot atomic vapor. The high optical depth (~1800) is the key for successful storage of a light with large single-photon detuning. The optical depth of our cold atomic ensemble is only about 38, which is too small to observe the clear storage of the idler. According our experimental conditions, $\Delta_{pr}=\Delta_c=0$MHz, $\Delta_p=\Delta_i=2\pi\times 80$MHz, $\gamma_{32}\approx\gamma_{31}=2\pi\times 3$MHz and $\gamma_2=0.008\gamma_{31}$ [34], we estimate that the ratio of intensities between the retrieved probe and the idler is about $10^4$:1.

After a certain storage time we switch on the coupling and pump beams again, then the spin state $\rho_{12}(z,t)$ converts back into optical mode. We should only observe the retrieved probe because the generated idler is hardly stored. But from the experimental data shown in Fig. 3, we could see the clear idler signal, besides probe signal. This is due to a new FWM process during the retrieval process: a new idler is generated through FWM among the retrieved probe field, the coupling field and the pump field. We also detect the beating signal with a beat frequency of 80MHz. (Fig. 3(c-d)).

In order to analyze the physical mechanism in this system, we perform the experiments in different situations. In situation 1, we turn pump and coupling fields on simultaneously in storing and retrieving processes. (Hereafter we use "writing" and "reading" instead of "storing" and "retrieving".) The obtained results shown in Fig. 4 (a) are same to Fig. 3 (a). In situation 2, we switch the pump field off and keep the coupling beam on in the writing process, but switch the pump and coupling on simultaneously in the reading process. Fig. 4 (b) shows the results, an idler signal is still obtained. In the writing process, there is no FWM process due to the absence of the pump field. So we think it is from the new FWM in the reading process: the retrieved probe field, the coupling field and the pump field combine to generate the new FWM signal. In situation 3, we switch the pump field off and keep the coupling beam on in the writing process. In the reading process, we switch the coupling off and keep the pump on. The results are obtained in Fig. 4 (c), where the idler signal is still observed. We think the observed idler signal come from the spontaneously Raman scattering. Because the coherence between ground states is established by the coexistence of coupling and probe beams previously, therefore the signal from spontaneously Raman scattering is amplified, so we can observe relative strong idler signal [36].

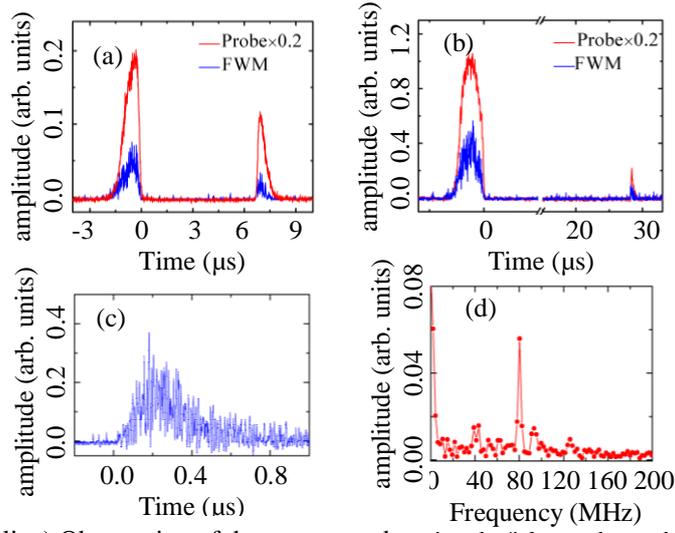

Fig. 3 (Color online) Observation of the storage and retrieval of the probe and FWM fields. (a) and (b) are storage and retrieval of the two pulses with a storage time of 6.7μs and 28μs respectively. (c) is the beating signal detected by PMT3 and in (d) the Fourier transform of the beating signal which shows an evident peak in the point of 80MHz.

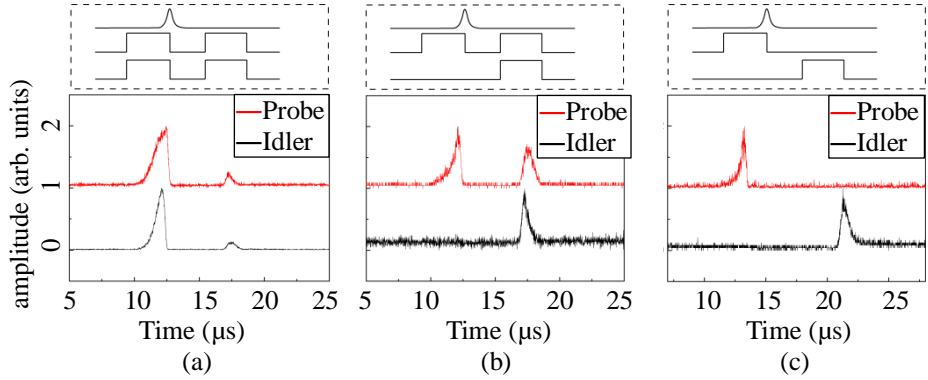

Fig. 4 (Color online) (up) The solid lines inside dotted rectangle are timing sequences of probe (top), coupling (middle) and pump (bottom) fields respectively. (down) The measured signals in three different situations. See the text for detail.

Next, we shift all the input beams +80MHz with respect to the formers, so now the probe and coupling fields are +80MHz detuning to the transition of $F=3\rightarrow F'=2$, and the pump field becomes resonant to the transition of F=3→F'=2. We redo the storage experiments as Fig. 4(a) and obtain the results shown in Fig. 5(a). In this condition, FWM condition is still satisfied. We do not observe the retrieved probe signal, this is because the probe field is difficult to be stored due to its +80MHz detuning. We can still observe the idler signal. We know that there is almost no probe field retrieved in the reading process, therefore generating the new FWM field in the reading process is impossible. So the retrieved idler field is from the atomic spin wave (i.e., stored FWM field generated in the writing process). Similarly, when we switch the pump field off in the writing process, and switch it on in the reading process, we could not observe the retrieved probe. The retrieved idler field is observable, but is very weak (almost zero). The results are shown in Fig. 5(b). Because there is no FWM field in both writing and reading processes, we think the observed idler is from the spontaneously Raman scattering. In addition, there is no coherence between ground states established before, so the observed signal is quite weak compared with the case of Fig. 4(c).

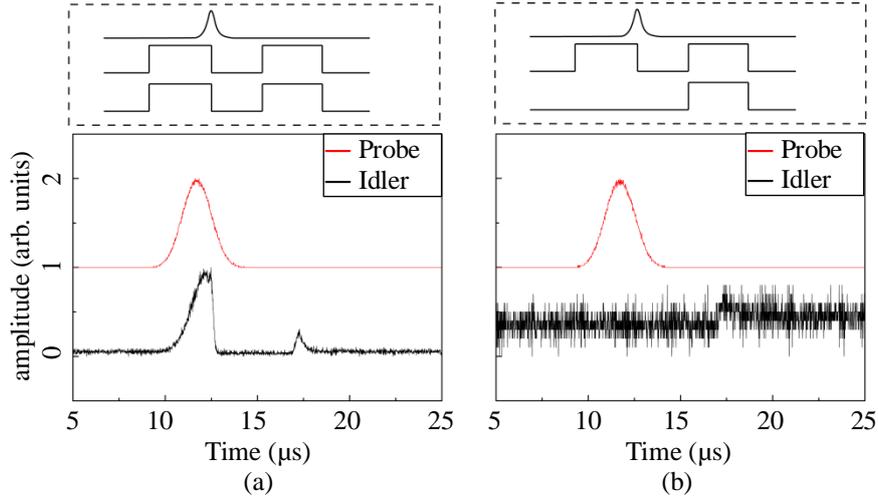

Fig. 5 (Color online) (up)The solid lines inside dotted rectangle are timing sequences of probe (top), coupling (middle) and pump (bottom) fields respectively. (down) The measured signals. See the text for detail.

According to the previous analysis, we see whether the probe and idler could be stored together in a FWM configuration strongly depends on the experimental condition, especially on the single-photon detuning of them. If they are both resonant to atomic transitions, then they could be stored simultaneously in an atomic ensemble. Vice-visa, they cannot be stored together.

We next apply an external homogeneous magnetic field generated by three pairs of Helmholtz coils (not shown) to atomic ensemble. In this experiment we choose the magnetic field orientation along the $z$-axis ($\theta=0$). During the storage process, the presence of such an external uniform magnetic field causes rotation of the atomic hyperfine coherences, leading to collapses and revivals oscillations of the DSPs. In the retrieving process the DSP will convert back to the optical mode. Detecting the retrieving optical mode at different storage time will reflect the phase rotation of the atomic spin excitation. The experimental conditions are the same as the Fig. 3. We input a probe pulse with a Gaussian shape in the presence of the coupling and pump beams, and the accompanying idler pulse is generated. We then adiabatically switch off the coupling and pump beams at the time when the peak of the Gaussian probe pulse enters the ensemble (see Fig.1). After a certain storage time, we switch on the coupling and pump beams again to retrieve the probe and idler pulses. As said before, because the single-photon detuning of the pump is quite large, therefore the idler field is mainly from the FWM during the retrieval process. We calculate the retrieval efficiency for the probe and the generation efficiency for the idler signals at different storage time, and show results in Fig.6. The measured oscillatory period is 2.7μs, corresponding to a magnetic field of 0.56 Gauss.

In a uniform magnetic field, revivals of the DSP should occur at times equal to $nT_L = n \times 2\pi\hbar / |g_F \mu_B \vec{B}|$, where $T_L$ is the Larmor period for the system and $n$ can be a multiple of half-integer [20], $\mu_B$ is the Bohr magneton, $g_F$ is the Landé $g$ factor for levels $F$. Unlike the equal atomic distribution of all the Zeeman sublevels in Ref. [20, 21], we initially pump the atoms to the states of $|5^2S_{1/2}$,

$F=3$, $m_F=-3\rangle$ and $|5^2S_{1/2}, F=3, m_F=-2\rangle$. The ground states in the DSPs are degenerate and have the same Landé $g$ factor $g_F$. So the phases of the two DSPs in our experiment will dynamically oscillate at the same frequency $f=2/T_L$, which agrees well with our experimental data.

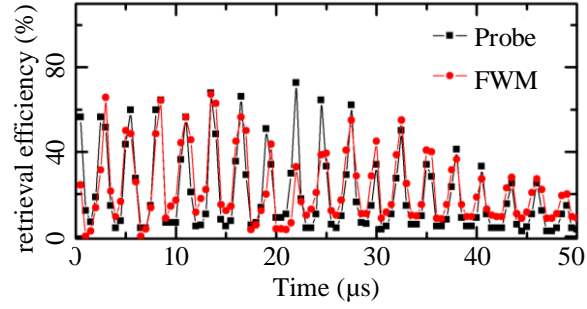

Fig. 6 (Color online) The collapse and revival oscillations of the retrieval efficiencies of the probe pulse (black) and the generated idler pulse (red).

## 4. Storage and retrieval of an image

Next, we report on the first experimental storage and retrieval of an image using a FWM process in this cold atomic ensemble. Instead of storing the image itself, we mapped the Fourier transform of the image [37, 38] into the atomic ensemble. This design can further reduce the diffusion of the atoms therefore helping us obtain the retrieved images with high fidelity. Besides, the usage of non-collinear backward-wave FWM geometry can significantly decrease the noises from the pump lasers.

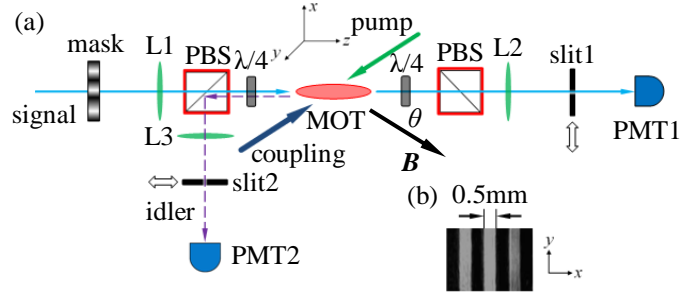

Fig. 7 (Color online) Experimental setup for the storage of an image. All the experimental laser beams come from one external-cavity diode laser. Their frequencies are modulated by some AOMs. The $1/e^2$ beam diameter of signal is approximately 5mm at the object plane and those of coupling and pump approximately 3.6mm and 3.2mm respectively. The peak powers of signal, coupling and pump are approximately 200pW, 200μW and 3.6mW respectively. Here, PBS is polarizing beam splitter; $\lambda/4$, quarter-wave plate; PMT, photomultiplier tube; mask, standard resolution chart (1951 USAF); L1~L3, lens. The transparent windows of slit1 and slit2 are aligned in the $y$-axis. (b) When we put a CCD camera in the position of slit1, the picture of cw signal imprinted with image is captured.

Fig. 7 shows the simplified experimental setup for the storage of the image. The energy levels are the same to Fig. 1. All beams are in the $x$-$z$ plane. Coupling and pump are collimated and coupled with each other. The angle between the signal (previous probe)-idler axis and coupling-pump axis is 1.5°. Signal comes out from a collimated fiber, then passes through a mask, and at last is focused to the center of atomic cloud (transform plane) by lens L1 in the *z*-axis. Coupling and pump totally cover signal in atomic cloud. We put the mask in the front focal plane of lens L1 (i.e., object plane). The front focal plane of lens L2 is coincident with the back focal plane of lens L1 in the transform plane. Hence, the object plane, lens L1, transform plane, lens L2 and image plane (the back focal plane of lens L2, where we will put CCD camera or slit1) compose a 4*f* imaging system. Because of the phase-matching condition in FWM process, the absolute value of the divergency of idler field is the same as that of the convergency of signal field. That means, another 4*f* imaging system consists of the object plane, lens L1, the transform plane, lens L3 and slit 2 (the back focal plane of lens L3). The image of signal after the 4*f* system is showed in Fig. 7(b).

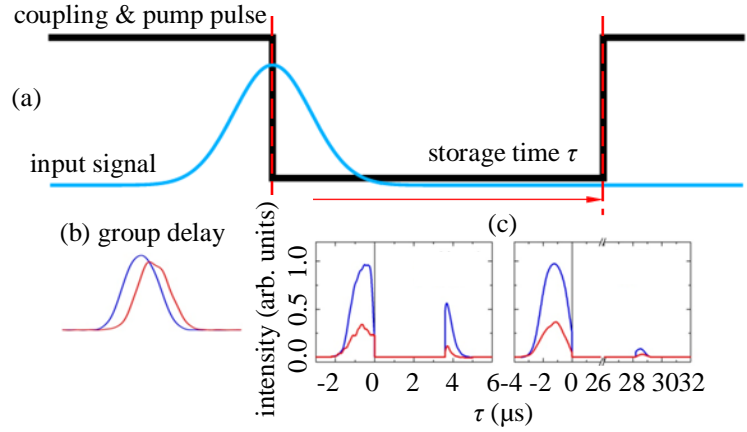

Fig. 8 (Color online) Representation of the synchronized timing of signal, coupling and pump. (b) Group delay of signal beams in our experiment is about 1μs. (c) Two experimental results of our signal (blue) and idler (red) storage and retrieval.

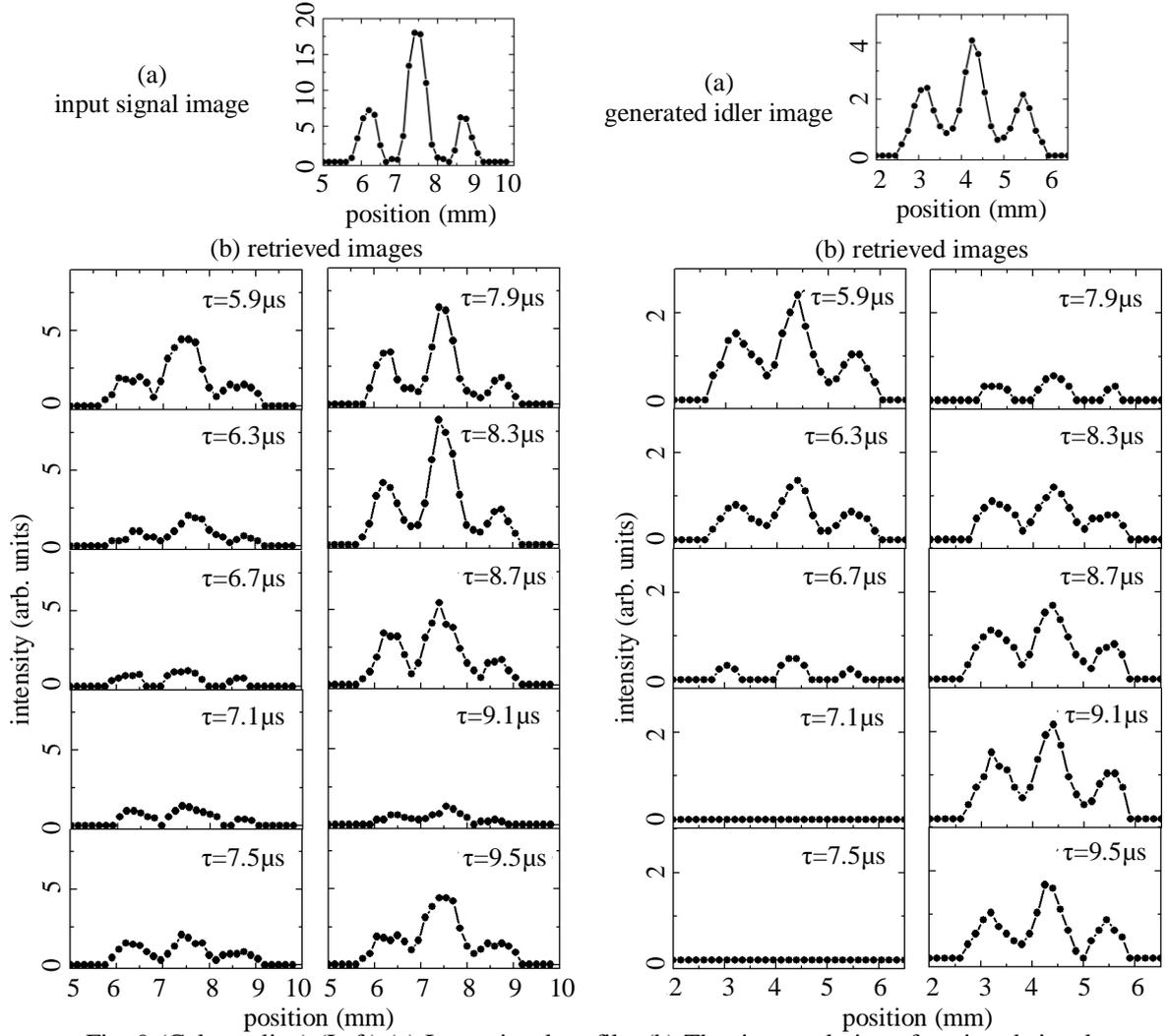

Fig. 9 (Color online) (Left) (a) Input signal profile. (b) The time evolution of retrieved signal fields. We observe the oscillation of retrieved signal versus storage time τ. (Right) (a) Generated idler profile. (b) The time evolution of retrieved idler fields. We observe the oscillation of retrieved idler versus storage time τ. The oscillating period is the same to that of signal.

The time sequence is showed in Fig. 8(a). Due to the steep dispersion of signal pulse, the group velocity will be greatly reduced (Fig. 8(b)). The delay time observed in our experiment is about 1μs (pump is shuttered and coupling is constant). In Fig. 8(c), we show two FWM storage and retrieval results. The full width at half maximum of the signal pulse is 1.3μs in our experiment. In the FWM process without storage, the images of signal and idler are shown in Fig. 9(left-a) and Fig. 10(Right-a), respectively. Here, the images information is extracted by scanning the slit1 in the $x$-axis and slit2 in the $z$-axis (Fig. 7(a)). The width of the slits is 0.4mm. The same method will be used in the extraction of retrieved images. Because there is no any spatial information carried in coupling and pump fields, the spatial dependence of idler comes from the signal field. Due to the phase-matching condition $k_s + k_i - (k_c + k_p) = 0$, where $k_s$, $k_i$, $k_c$ and $k_p$ are the wave-vectors of signal, idler, coupling and pump respectively, the generated idler field will also carry this spatial information.

Assuming the transmission function of the mask is $T(x) = t^2(x)$, and the transverse profile of signal before mask is Gaussian profile $I_{s0}\exp(-x^2/d^2)$, where $I_{s0}$ is the peak intensity and $d$ the $1/e^2$ diameter, the profile obtained at object plane is $I_s(x) = T(x) \cdot I_{s0}\exp(-x^2/d^2)$, and the profile of signal after the slit1 is indicated in Fig. 7(a). Furthermore, the vectorial amplitude of signal is square-root of $I_s(x)$, that is, $\mathbf{E}_s(x) = t(x) \cdot \mathbf{E}_{s0}\exp(-x^2/2d^2)$. At the transform plane, the amplitude of signal $\mathbf{E}_{sf}(\xi)$ is given by the Fourier transform [38]

$$\mathbf{E}_{sf}(\xi) = \mathcal{F}(\mathbf{E}_s) = \frac{1}{\sqrt{\lambda f}} \int_{-\infty}^{\infty} \mathbf{E}_s(x) \exp(-i\frac{2\pi}{\lambda f} x\xi) dx, \quad (6)$$

where $\xi$ is the coordinate of Fourier transform plane, $\lambda$ the center wavelength of signal and $f$ the focal length of lens L1. The amplitude of idler at the transform plane is $\mathbf{E}_{if}(\xi)$. The spatial pattern and direction of $\mathbf{E}_{if}(\xi)$ are totally determined by the phase-matching condition $\mathbf{k}_s + \mathbf{k}_i - (\mathbf{k}_c + \mathbf{k}_p) = \mathbf{0}$. The vectorial ground state coherence during the storage is given by [28, 33]

$$\boldsymbol{\rho}_{12}(t,\xi) = \boldsymbol{\rho}_{12}(0,\xi)\exp(-\frac{t}{2\tau}). \quad (7)$$

Here, the diffusion of the atoms is neglected. The time constant $\tau$ shows an exponential decay of the energy. As shown in Fig. 8(a), when we adiabatically turn off the coupling and pump, the signal is mapped into the long-lived ground state coherence $\boldsymbol{\rho}_{12}$. After some time, we turn coupling and pump on, and $\boldsymbol{\rho}_{12}$ will be converted back into the electromagnetic fields, i.e., signal. The idler field with image is also retrieved due to the reconstruction of FWM process. The spatial patterns of retrieved signal and idler are completely determined by the spatial pattern of $\boldsymbol{\rho}_{12}$ as we see from Eq. 7.

The backward-wave geometry and the non-collinear configuration between signal axis and pump axis make storage in single-photon level possible. Of course, our scheme could be definitely applied to a hot atomic ensemble. The atoms in a cold ensemble move more slowly, thus the dephasing rate caused by atom motion is reduced significantly, which makes longer storage time possible compared to a hot atomic ensemble, Besides, because of the low temperature of atoms in MOT, the diffusion can be neglected, which enhances the fidelity of the retrieved images. The optical depth would greatly affect the efficiency of the storage and retrieval processes, and the fidelity of the retrieved images. Even though, the optical depth does not become too large, otherwise the quality of stored image would be compromised [5]. The detail discussions of differences with different ensembles are worth doing further in the future. Anyway, the storage and retrieval of an image will find applications in the quantum information processing, remote sensing and image processing.

## 5. Conclusion

To sum up, we perform several experiments of the storage and retrieval of the probe field and the idler field via an off-axis FWM process in a cold atomic ensemble. We analyze the underlying physical mechanism in these experiments. Our results clearly show whether the probe or idler generated through FWM could be stored depends on the atomic configuration. The results are very helpful to understand the physical mechanism of storage by a FWM process. Besides, we also report on the first experimental demonstration of the storage and retrieval of an image in a cold atomic ensemble using a FWM process. This image storage may have important implications in image processing, remote sensing and quantum communication, where two correlated fields need to be preserved for later use.

**Acknowledgments**

[1]These authors contributed to this work equally. Y. Liu would like to thank Prof. Jianming Wen for his helpful discussion. This work was supported by the National Natural Science Foundation of China (Grant Nos. 11174271, 61275115, 10874171) and the National Fundamental Research Program of China (Grant No. 2011CB00200).